\documentclass[twocolumn]{webofc}
\usepackage[varg]{txfonts}   
\usepackage{siunitx}
\usepackage{graphicx}
\usepackage[dvipsnames]{xcolor}
\usepackage{tikz}
\usepackage[normalem]{ulem}
\usepackage{ifthen}
\usetikzlibrary{arrows}
\usetikzlibrary{positioning}

\graphicspath{{images/}}

\newcommand{\perkeo}{\textsc{Perkeo~III}}
\newcommand{\figref}[1]{Figure \ref{#1}}
\newcommand{\punkt}{\mathrm{.}}
\newcommand{\komma}{\mathrm{,}}

\begin{document}

\title{Undetected Electron Backscattering in \perkeo{}}

\author{\firstname{Christoph} \lastname{Roick}\inst{1}\fnsep\thanks{\email{christoph.roick@tum.de}} \and
        \firstname{Heiko} \lastname{Saul}\inst{1,2} \and
        \firstname{Hartmut} \lastname{Abele}\inst{2} \and
        \firstname{Bastian} \lastname{Märkisch}\inst{1}\fnsep\thanks{\email{maerkisch@ph.tum.de}}
}

\institute{Physik-Department ENE, Technische Universität München, James-Franck-Str. 1, 85748 Garching, Germany
\and
           Atominstitut, Technische Universität Wien, Stadionallee 2, 1020 Wien, Austria
          }

\abstract{%
  The beta asymmetry in neutron beta decay is used to determine the ratio of axial-vector coupling to vector coupling most precisely. In electron spectroscopy, backscattering of electrons from detectors can be a major source of systematic error. We present the determination of the correction for undetected backscattering for electron detection with the instrument \perkeo{}. For the electron asymmetry, undetected backscattering leads to a fractional correction of $5\times 10^{-4}$, i.e. a change by \SI{40}{\percent} of the total systematic uncertainty.
}
\maketitle
\section{Introduction}
\label{intro}
Within the standard model of particle physics neutron beta decay is completely determined by three parameters only. These are the element $V_\mathrm{ud}$ of the Cabibbo-Kobayashi-Maskawa (CKM) quark mixing matrix,
the ratio between axial-vector and vector coupling constants $\lambda$ and the Fermi coupling constant $G_\mathrm{F}$. The latter is known very precisely from muon decay \cite{MuLanCollaborationMeasurementPositiveMuon2011}. The currently most precise observable to determine $\lambda$ is the beta asymmetry $A$ in polarized neutron decay \cite{MundDeterminationWeakAxial2013,UCNACollaborationPrecisionmeasurementneutron2013,MarkischMeasurementWeakAxialVector}. $V_\textup{ud}$ is most precisely determined from superallowed beta decays, but may also be extracted from $\lambda$ and the neutron lifetime with competitive precision \cite{darkchannel}. Inconsistencies in this overdetermined set of observables may hint to physics beyond the Standard Model \cite{darkchannel}.
 
\begin{figure}[h]
\centering
\includegraphics[width=\columnwidth]{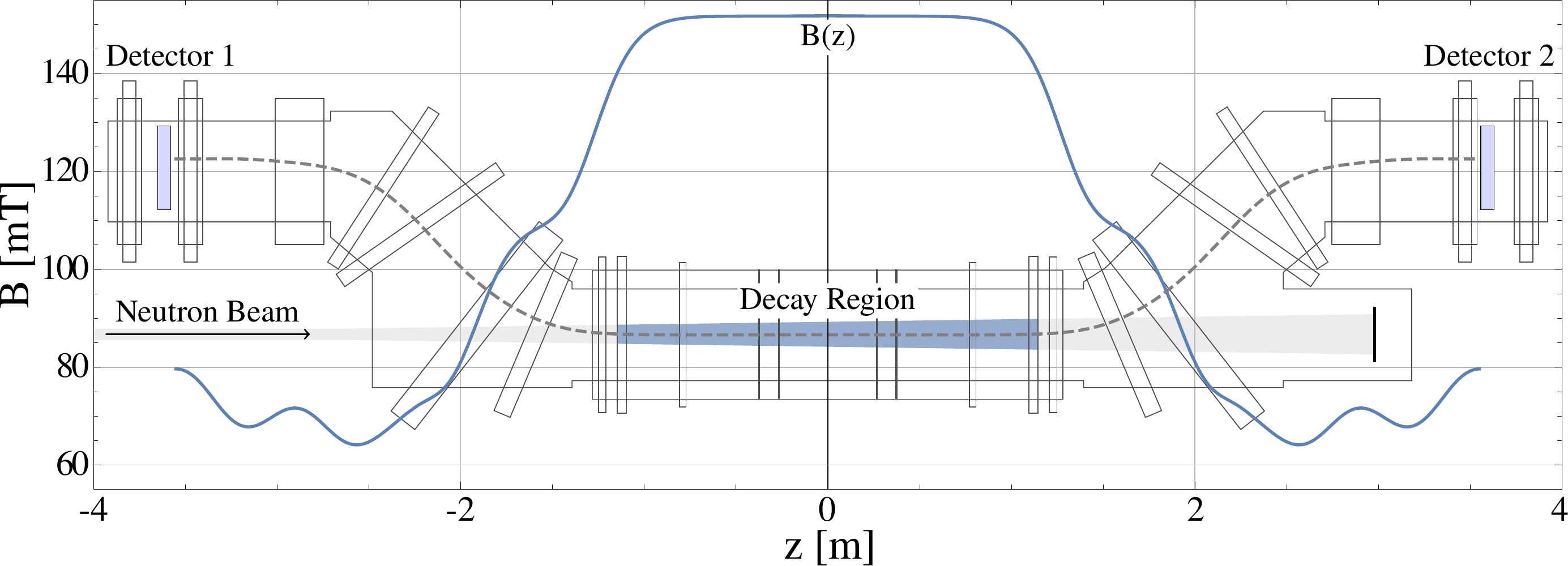}
\caption{Electrons originating from the central region ($|z|<\SI{1}{m}$) of \perkeo{} follow the magnetic field lines (dashed) and end up in one of the detectors located at $|z|\approx\SI{3.6}{m}$. Due to the inverse magnetic mirror effect, the maximum polar angle of the electron momentum is $\approx\SI{47}{\degree}$ in the detector region.}
\label{fig:perkeo}
\end{figure}

In the case of the neutron decay spectrometer \perkeo{} \cite{Markischnewneutrondecay2009} , $\lambda$ is obtained from a determination of the experimental electron asymmetry
\begin{align}
  A_\textup{exp}=\frac{N_\uparrow - N_\downarrow}{N_\uparrow + N_\downarrow}=\frac{1}{2}A\beta(E_e) P\komma
\end{align}
where $N_\uparrow$ ($N_\downarrow$) denotes the number of electrons emitted in (against) the direction of neutron polarization $P$. Within the Standard Model of particle physics, the electron asymmetry parameter $A$ is expressed by
\begin{align}
  A = -2\frac{\lambda+\lambda^2}{1+3\lambda^2}\komma
\end{align}
neglecting small correction terms \cite{WilkinsonAnalysisneutronvdecay1982,IvanovNeutronBetaDecayLaboratory2013}.
In \textsc{Perkeo} the sign of $P$ is switched regularly using an adiabatic fast passage spin flipper with an efficiency close to unity.
Electrons from the decay region are magnetically guided to scintillation detectors. \figref{fig:perkeo} shows the configuration of the magnetic field. The setup allows for a $4\pi$ solid angle coverage of electron detection for decays which happen in the central region.
Energy reconstruction is done by photon counting of the scintillation light by employing mesh photomultiplier tubes (PMTs) which are coupled to the scintillator using light guides. About $10^3$ events/s are measured while the neutron pulse is in the central region.

In this work we present a correction to an effect which arises from the combination of electron backscattering from the electron detectors and a limited trigger efficiency. This undetected backscattering may lead to an erroneous assignment of electron emission to the wrong hemisphere and an insufficient energy reconstruction.

\subsection{Backscattering}
\label{backscattering}
In electron backscattering the electron only deposits a fraction of its full energy in the scintillator before leaving the detector again. In order to guarantee full energy reconstruction even for backscattered electrons, in \perkeo{} both detectors are connected by the magnetic field. About one half of the backscattered electrons can overcome the magnetic barrier of the central magnetic field and are guided to the opposite detector. The other half will be reflected by the magnetic field and hit the original detector again.

\begin{figure}[h]
\centering
\includegraphics[width=0.95\columnwidth]{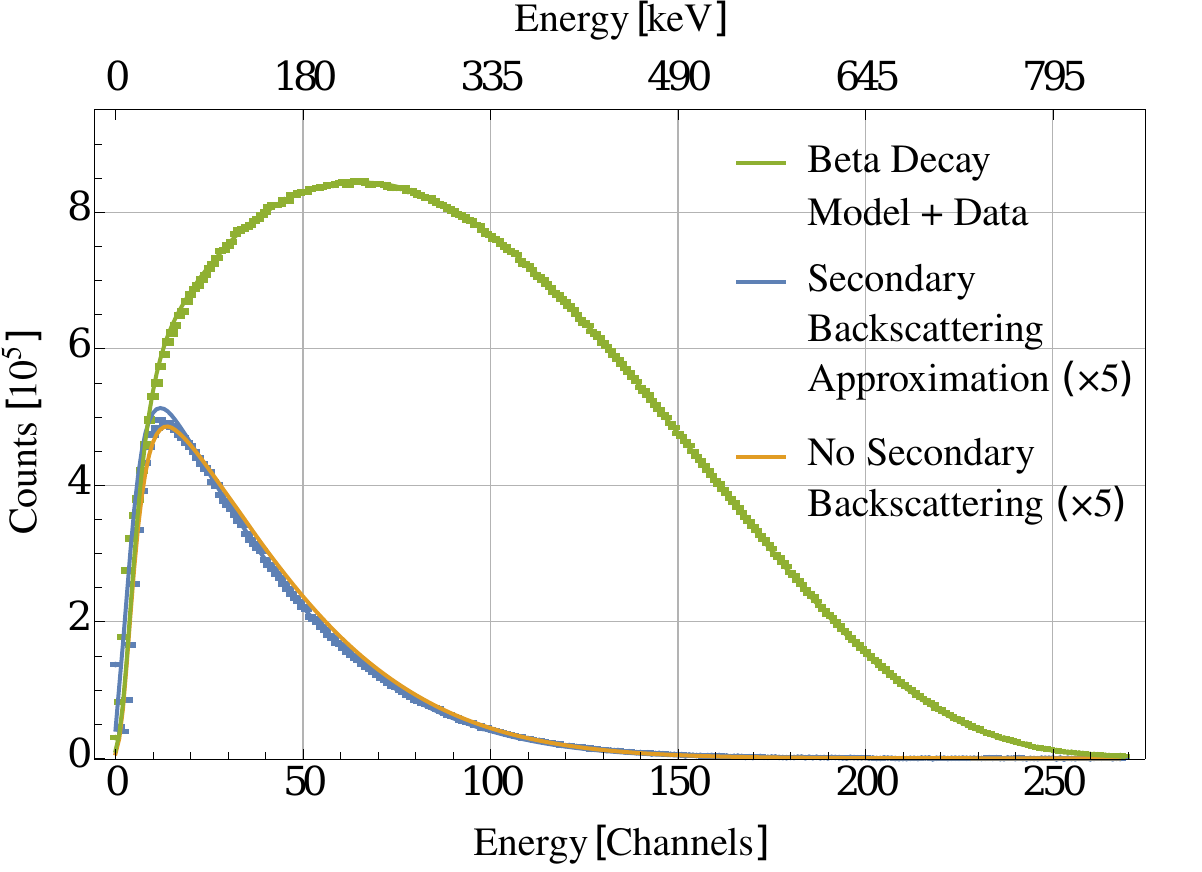}
\caption{A simulated backscattering spectrum (scaled by a factor of~\num{5}) is compared to data from neutron decay measurements with \perkeo{}. If considering backscattering of backscattered electrons -- denoted as secondary backscattering -- simulation (solid lines) and data agree well. The norm is obtained from the complete beta decay spectrum, all detector parameters are fixed by measurements with calibration sources.}
\label{fig:backscattering}
\end{figure}

An estimate for the probability and energy and angular dependence of electron backscattering is obtained from simulations using \texttt{Geant4 v10.3.1} \cite{AllisonRecentdevelopmentsGeant42016}. For low-energy scattering simulations, the physics models have to be selected with care. We follow suggestions by \cite{KimValidationTestGeant42015}. According to the simulations, the probability for backscattering to the opposite detector in \perkeo{} is about \SI{6}{\percent}. This result agrees well with measured backscattering spectra in neutron decay, see \figref{fig:backscattering}.

\subsection{Trigger Function}
The trigger function $T(L)$, which will be presented in more detail in \cite{SaulPMTPaper2019}, generates the actual measured spectrum to a number of photoelectrons $L$ with trigger thresholds applied. In general it depends also on $L$ and not only on the theoretical detector response $S(L)$ -- the measured signal spectrum without trigger threshold. The description of the trigger function includes the electronic trigger threshold, coincidence conditions which are imposed to reduce the registration of PMT dark counts, the response function of PMTs to a number of initial photoelectrons, and the distribution of the emitted light $L$ onto different light guides, depending on the position of electron impact on the detector. With this we determine the trigger function from the measured signal of a point-like calibration source and subsequently apply it to the extended source of the neutron beam.

\subsection{Further Detector Parameters}
We consider the point spread \cite{psf1, psf2} of the electrons as they are mapped from their source onto the detector. This spread is convoluted with the spatial non-uniformity of the detector, which arises from differences in the efficiency of light transport to the PMTs, depending on the position on the light origin.

Furthermore, there are non-linear effects for one during the light creation due to quenching \cite{Birks} and also in the analogue part of the data acquisition system. We consider these effects with an effective overall non-linearity that is part of the light creation process and therefore entirely included in $L$. Only in future experiments this approximation might be handled differently. Finally, there are broadening effects due to the finite number of photoelectrons, the amplification process in the PMTs and electronic noise \cite{RoickParticleDetectionProton2018, ThesisSaul}.

\section{Undetected Backscattering}
\label{sec:ubs}
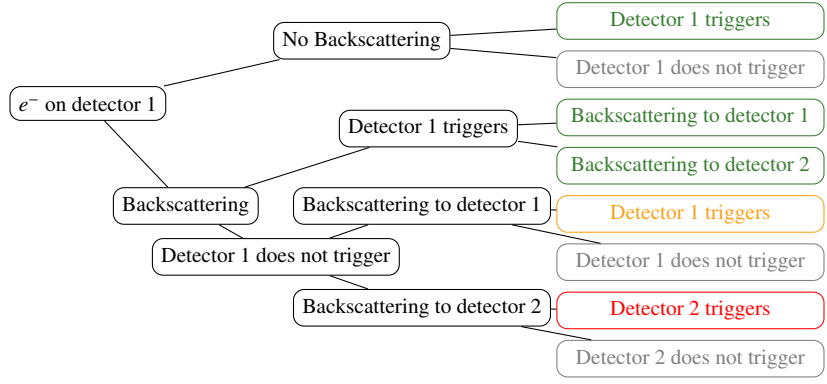
\begin{figure*}[h]
\centering
\begin{tikzpicture}[grow = right, every node/.style = {shape=rectangle, rounded corners, draw, align=center, font=\footnotesize}, last/.style = {minimum width=10em}]
  \node (init) {$e^-$ on detector 1}
    child { node [below right = 2.5em and -2em of init] (bs) {Backscattering}
      child { node [below right = 0.5em and -4em of bs] (bsnt) {Detector 1 does not trigger}
        child { node [below right = 0.5em and -4em of bsnt] (bsnt2) {Backscattering to detector 2}
          child { node [last, below right = 0.5em and 0.2em of bsnt2, color = gray] (bsnt2n) {Detector 2 does not trigger} }
          child { node [last, above = 0.4em of bsnt2n, color = red] (bsnt2t) {Detector 2 triggers} }
        }
        child { node [above right = 0.5em and -4em of bsnt] (bsnt1) {Backscattering to detector 1}
          child { node [last, above = 0.4em of bsnt2t, color = gray] (bsnt1n) {Detector 1 does not trigger} }
          child { node [last, above = 0.4em of bsnt1n, color = YellowOrange] (bsnt1t) {Detector 1 triggers} }
        }
      }
      child { node [above right = 1.5em and 3em of bs] (bst) {Detector 1 triggers}
        child { node [last, above = 0.4em of bsnt1t, color = OliveGreen] (bst2) {Backscattering to detector 2} }
        child { node [last, above = 0.4em of bst2, color = OliveGreen] (bst1) {Backscattering to detector 1} }
      }
    }
    child { node [above right = 1em and 4em of init] (nobs) {No Backscattering}
      child { node [last, above = 0.4em of bst1, color = gray] (nobsnt) {Detector 1 does not trigger} }
      child { node [last, above = 0.4em of nobsnt, color = OliveGreen] {Detector 1 triggers} }
    };
\end{tikzpicture}
\caption{A first order decision tree is used to visualize the effects of backscattering. \textcolor{OliveGreen}{Green} boxes denote events with correct emission direction and energy information. \textcolor{gray}{Gray} boxes denote events without any trigger at all, while \textcolor{YellowOrange}{yellow} stands for the right emission direction assignment, but detection at lower energy, and \textcolor{red}{red} misses both, emission direction and full energy reconstruction.}
\label{fig:bs_decision_tree}
\end{figure*}

The combination of electron backscattering and the trigger function leads to an important systematic effect for the determination of $A_\textup{exp}$ \cite{RoickParticleDetectionProton2018}. Several cases have to be considered if backscattering occurs. A backscattered electron may either be transported to the opposite detector if the emission angle to the normal of the detector surface $\theta < \theta_\textup{lim}$ or be reflected on the magnetic field for $\theta > \theta_\textup{lim}$, impinging on the original detector again. The limiting angle $\theta_\textup{lim} = \arcsin\sqrt{B_\textup{detector}/B_{z=0}}\approx \SI{47}{\degree}$ in \perkeo{} is defined by the magnetic mirror effect. Typical electron flight times between detectors are on the order of a few $\SI{10}{ns}$.
\figref{fig:bs_decision_tree} visualizes possible event chains including up to one backscattered electron. 

For the measurement of the electron asymmetry, the initial impact is crucial as it leads to an assignment of the event to $N_\uparrow$ ($N_\downarrow$) -- emission in (against) neutron spin direction.
The trigger function plays an important role, since the deposited energy in the primary detector might not be sufficient to release a trigger signal.
In this case with backscattering towards the opposite detector, the event could be assigned to the wrong emission direction, if the deposited energy is sufficiently large for a trigger there instead.

To account for non-linear effects, we introduce a response function $L(E)$ describing the photoelectron response to an electron that completely deposits its energy inside one of the scintillators. Potential dead layers have negligible influence on the spectrum. In general $L(E)$ are normalized spectra of photoelectrons. For the sake of readability, they will be considered numbers in the following. If an electron gets backscattered from the first detector with remaining energy $E_2$, its photoelectron response is $L_1 = L(E) - L(E_2)$ in the first detector and $L_2 = L(E_2)$ in the second detector. Due to the non-linearity of the detector response, $L(E) - L(E_2) \ne L(E - E_2)$. For a backscattering event, the probability to correctly measure the full event is%
\begin{align}
 p_\textup{Tbs} = |T(L_1)| = |T(L(E) - L(E_2))|\komma
\end{align}
with the trigger function $T(L)$. Bars $|T|$ denote the $\mathcal{L}^1$-norm of a spectrum $T$.
In this case it is irrelevant whether the second impact releases a trigger as well, as the data acquisition is always running for both detectors simultaneously during a reasonable time. The remaining number of photoelectrons $L(E_2)$ is therefore correctly assigned to the corresponding event.

In case of undetected backscattering, i.e. if the first impact does not release a trigger, but the second impact is registered, the probability is expressed as
\begin{align}
 p_\textup{Tubs} = (1 - |T(L_1)|) \cdot |T(L_2)|\punkt
\end{align}
Assuming two identical detectors, the probability distribution of detected energy $E_d$ in both detectors of an initial electron with energy $E$ now is
\begin{align}
\begin{aligned}
  \frac{\textup{d}I_T}{\textup{d}E_d}&\bigg\rvert_E=
  [\text{undetected backscattering}]\\
   &+\left([\text{no backscattering}]+[\text{detected backscattering}]\right)\\
  =&\left[\eta_\textup{same}(E,E_d) + \eta_\textup{opp}(E,E_d)\right]\cdot\\
   &\quad\underbrace{T(L(E_d)) \cdot \left[1 - |T(L(E) - L(E_d))|\right]}_{\substack{\text{\footnotesize{\hspace{3.5em}spectrum/probability of}}}}\\
   &+\Bigg[1 - \int_0^E\bigg(\overbrace{|T(L(E_2))| \cdot \left[1 - |T(L(E) - L(E_2))|\right]}^{\substack{\text{\footnotesize{\hspace{-1.8em}normalized undetected backscattering}}}}\\
   &\phantom{+\Bigg(1 - \int_0^E}\cdot \overbrace{(\eta_\textup{same}(E, E_2) + \eta_\textup{opp}(E, E_2))}^{\substack{\text{\footnotesize{backscatter probability}}}}\bigg)\,\mathrm{d}E_2\Bigg]\cdot\\
   &\quad\delta(E_d - E)\komma
\end{aligned}
\end{align}
with $\eta_\textup{opp/same}(E, E_2)$ being the probability of backscattering with incident energy $E$ and energy of the backscattered electron $E_2$ to the opposite or the same detector respectively.

\subsection{Impact on the Measured Asymmetry}
Assuming a negligible trigger threshold the measured asymmetry including backscattering is hence described by
\begin{align}
 \begin{split}
 A_\textup{exp}&=\frac{N_\uparrow-N_\downarrow}{N_\uparrow+N_\downarrow}\\
 N_\uparrow&=N^\textup{nbs}_\uparrow + D^\textup{same}_\uparrow + D^\textup{opp}_\uparrow + U^\textup{same}_\uparrow + \boldsymbol{U^\textup{opp}_\uparrow}\\
 N_\downarrow&=N^\textup{nbs}_\downarrow + D^\textup{same}_\downarrow + D^\textup{opp}_\downarrow + U^\textup{same}_\downarrow + \boldsymbol{U^\textup{opp}_\downarrow}\komma
 \end{split}
 \label{eq:asymnormal}
\end{align}
where $N^\textup{nbs}$ is used for the spectrum without backscattering, $D^\textup{same}$ and $D^\textup{opp}$ for the spectra of detected backscattering onto the same or the opposite detector and $U^\textup{same}$ and $U^\textup{opp}$ for the respective undetected spectra, namely the triggered spectrum of the remaining energy after backscattering. The experimental asymmetry including the finite detection threshold is now
\begin{align}
 \begin{split}
 A_\textup{exp}'&=\frac{N'_\uparrow-N'_\downarrow}{N'_\uparrow+N'_\downarrow}\\
 N'_\uparrow&=N^\textup{nbs}_\uparrow + D^\textup{same}_\uparrow + D^\textup{opp}_\uparrow + U^\textup{same}_\uparrow + \boldsymbol{U^\textup{opp}_\downarrow}\\
 N'_\downarrow&=N^\textup{nbs}_\downarrow + D^\textup{same}_\downarrow + D^\textup{opp}_\downarrow + U^\textup{same}_\downarrow + \boldsymbol{U^\textup{opp}_\uparrow}\komma
 \end{split}
 \label{eq:asymcorrected}
\end{align}
where we note that the component $U^\textup{opp}$ now carries the \emph{other} spin component and therefore affects the measured asymmetry. Assuming equal detector response functions, eq.~\eqref{eq:asymcorrected} can for now be simplified to
\begin{align}
 \begin{split}
 A_\textup{exp}' &\approx \frac{(\alpha-\epsilon)N_\uparrow + \epsilon N_\downarrow - (\alpha-\epsilon)N_\downarrow - \epsilon N_\uparrow}{(\alpha-\epsilon)N_\uparrow + \epsilon N_\downarrow + (\alpha-\epsilon)N_\downarrow + \epsilon N_\uparrow}\\
 &= \left(1-2\frac{\epsilon}{\alpha}\right)\frac{N_\uparrow - N_\downarrow}{N_\uparrow + N_\downarrow}=\left(1-2\frac{\epsilon}{\alpha}\right)A_\textup{exp}\komma
 \end{split}
\end{align}
with the triggered fraction without backscattering and detected backscattering $\alpha(E)$ and the fraction of undetected backscattering onto the opposite detector $\epsilon(E)$ for clean and untriggered spectra $N$. Here we assume that the energy can fully be reconstructed, even in the case of undetected backscattering.

We note that a previous analysis \cite{SchumannUnrecognizedBackscatteringLow2008} underestimates $\epsilon/\alpha$, mainly due to oversimplified assumptions about the detector function when extrapolating the effect from the data. For the determination of the electron asymmetry parameter with \textsc{Perkeo~II} \cite{MundDeterminationWeakAxial2013}, the resulting correction was already increased in order to comply with a reevaluated detector function.

\subsection{Unequal Detector Functions}
In general the detector response functions $S_{1(2)}(L)$ -- the theoretical detector response without trigger condition -- and trigger functions $T_{1(2)}(L)$ are different for both detectors. Furthermore, we cannot assume that the data acquisition is started early enough to allow for the reconstruction of the first energy deposition in the case of undetected backscattering: The flight time of an electron between first and second hit is probably too long to register the untriggered detector response to the first energy deposition. These constraints are considered in eq. \eqref{eq:ubsnoreconstruct}, where the measured spectrum of electrons coming from neutrons which are polarized towards detector \num{1} is represented.
$E_1$ denotes the full energy of events which include undetected backscattering, whereas $E_2$ is the energy of the backscattered electrons, which already deposited sufficient energy for a trigger signal after the first hit. \figref{fig:nup} shows the individual components to classify their impact on the measured spectrum.

\subsection{Estimation of Systematic Effects}
An important part of the correction is the determination of the actual trigger function. We use both methods described in \cite{SaulPMTPaper2019} and \cite{SchumannUnrecognizedBackscatteringLow2008} for its characterization and obtain agreement on the order of \SI{2}{\percent}. Together with the uncertainties for the description of the underlying theoretical spectra and timing effects on the ADC signal we estimate the uncertainty to \SI{5}{\percent}.

Further uncertainties are the calculation of the backscattering probabilities $\eta_\textup{same}$ and $\eta_\textup{opp}$, where the latter has the main contribution to the correction. The magnetic field in the detector region for this measurement is known only on a \SI{10}{\percent} level, which hardly changes $\eta_\textup{same} + \eta_\textup{opp}$, but shifts between $\eta_\textup{same}$ and $\eta_\textup{opp}$ due to a change of the limiting angle $\theta_\textup{lim}$ of the magnetic mirror effect. Simplifying the angular distribution of backscattered electrons to $1-(4\theta/\pi-1)^2\approx\sin2\theta$ \cite{Wietfeldtbackscattersuppressedbetaspectrometer2005} results in
\begin{align}
  \frac{\eta_\textup{opp}}{\eta_\textup{same} + \eta_\textup{opp}} &= \cos^2\sqrt{B_1/B_0}\\
  \Delta\frac{\eta_\textup{opp}}{\eta_\textup{same} + \eta_\textup{opp}} &= \frac12 \sqrt{\frac{B_0}{B_1}} \sin\left(2\sqrt{B_1/B_0}\right)\Delta\frac{B_1}{B_0} \approx \SI{7}{\percent}\komma
\end{align}
which is increased to \SI{10}{\percent} to account for the simplified angular distribution of backscattered electrons.
To account for the neglect of secondary backscattering, the choice of the ionization model and uncertainties in the angular distribution of the scattering model, an uncertainty of \SI{20}{\percent} -- see also \cite{MartinNewmeasurementsquantitative2006} -- is assumed for the determination of energy dependent backscattering coefficients from simulations.

A model dependent uncertainty is the reconstruction of the energy of undetected events. The ADC signal registration starts a few \SI{}{ns} before the triggering signal enters the data acquisition system. If the time of flight for an undetected backscattered electron is smaller than this timing buffer, then the energy is reconstructed, otherwise it is lost or partly lost. The effect is maximized if full energy reconstruction is assumed for all events. It changes the addition of the incompletely reconstructed energy of eq. \eqref{eq:ubsnoreconstruct}
\begin{align}
\int\limits_E^{E_\textup{max}}\!\!\!\!\eta(E_1, E)\boldsymbol{N_\downarrow(E_1)}\cdot (1-|T_2(L(E_1)-L(E))|)\cdot T_1(L(E))\,\mathrm{d}E_1
\end{align}
to a convolution of the untriggered spectrum on the detector first hit with the triggered spectrum for the parts of undetected backscattering:
\begin{align}
\int\limits_0^E\!\!\!\eta(E, E_2)N_\downarrow(E)\cdot [1-T_2(L(E)-L(E_2))]\boldsymbol{*} T_1(L(E_2))\,\mathrm{d}E_2\punkt
\end{align}
The correction including full energy reconstruction increases the correction by \SI{15}{\percent}. From the timing of the electronics we assume that energy reconstruction is (typically) not possible and consider the difference to be an uncertainty.
\begin{figure*}[h]
\centering
\begin{align}
 \begin{aligned}
 N'_\uparrow(E)=
 &(1-\eta_\textup{same}(E)-\eta_\textup{opp}(E))N_\uparrow(E)\cdot T_1(L(E)) &\text{\footnotesize{no backscattering\,(bs)} }&\, N^\textup{nbs}_\uparrow\\
 &+\int\limits_0^E\eta_\textup{same}(E, E_2)N_\uparrow(E) \cdot \left[T_1(L(E)-L(E_2))*S_1(L(E_2))\right]\,\mathrm{d}E_2 &\text{\footnotesize{detected bs to D1} }&\, D^\textup{same}_\uparrow\\
 &+\int\limits_0^E\eta_\textup{opp}(E, E_2)N_\uparrow(E)\cdot \left[T_1(L(E)-L(E_2))*S_2(L(E_2))\right]\mathrm{d}E_2 &\text{\footnotesize{detected bs to D2} }&\, D^\textup{opp}_\uparrow\\
 &+\int\limits_E^{E_\textup{max}}\eta_\textup{same}(E_1, E)N_\uparrow(E_1)\cdot (1-|T_1(L(E_1)-L(E))|)\cdot T_1(L(E))\,\mathrm{d}E_1 &\text{\footnotesize{undetected bs from D1} }&\, U^\textup{same}_\uparrow\\
 &+\int\limits_E^{E_\textup{max}}\eta_\textup{opp}(E_1, E)\boldsymbol{N_\downarrow(E_1)}\cdot (1-|T_2(L(E_1)-L(E))|)\cdot T_1(L(E))\,\mathrm{d}E_1 &\text{\footnotesize{undetected bs from D2} }&\, U^\textup{opp}_\downarrow
 \end{aligned}
 \label{eq:ubsnoreconstruct}
\end{align}
\end{figure*}
\begin{figure}
\centering
\includegraphics[width=0.95\columnwidth]{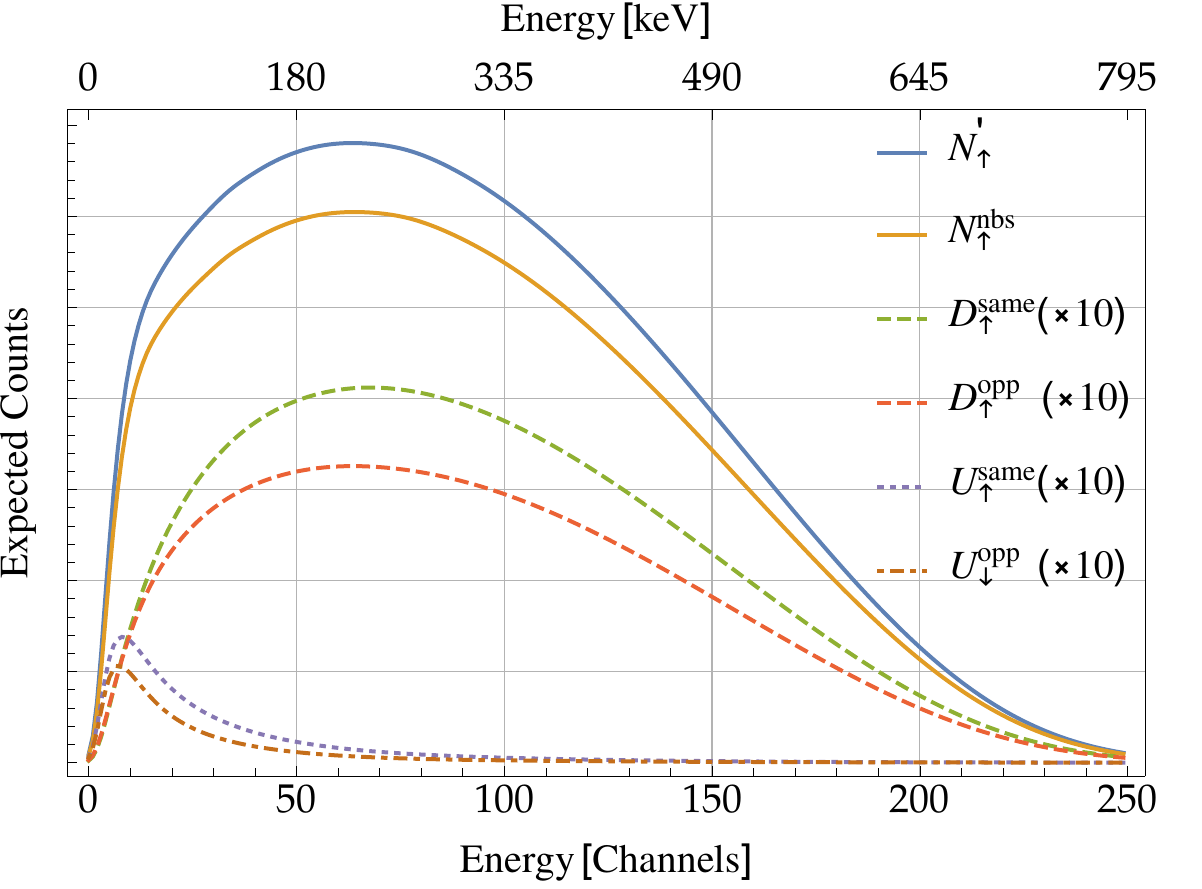}
\caption{The individual components of the corrected number of counts $N'_\uparrow$ from eq. \eqref{eq:asymcorrected}, as calculated in eq. \eqref{eq:ubsnoreconstruct}. Spectra including backscattering are enhanced by a factor of \num{10}. Undetected backscatter events $U^\textup{opp}_\downarrow$ from the opposite detector are assigned to the wrong hemisphere.}
\label{fig:nup}
\end{figure}

\begin{figure}[h]
\centering
\includegraphics[width=0.95\columnwidth]{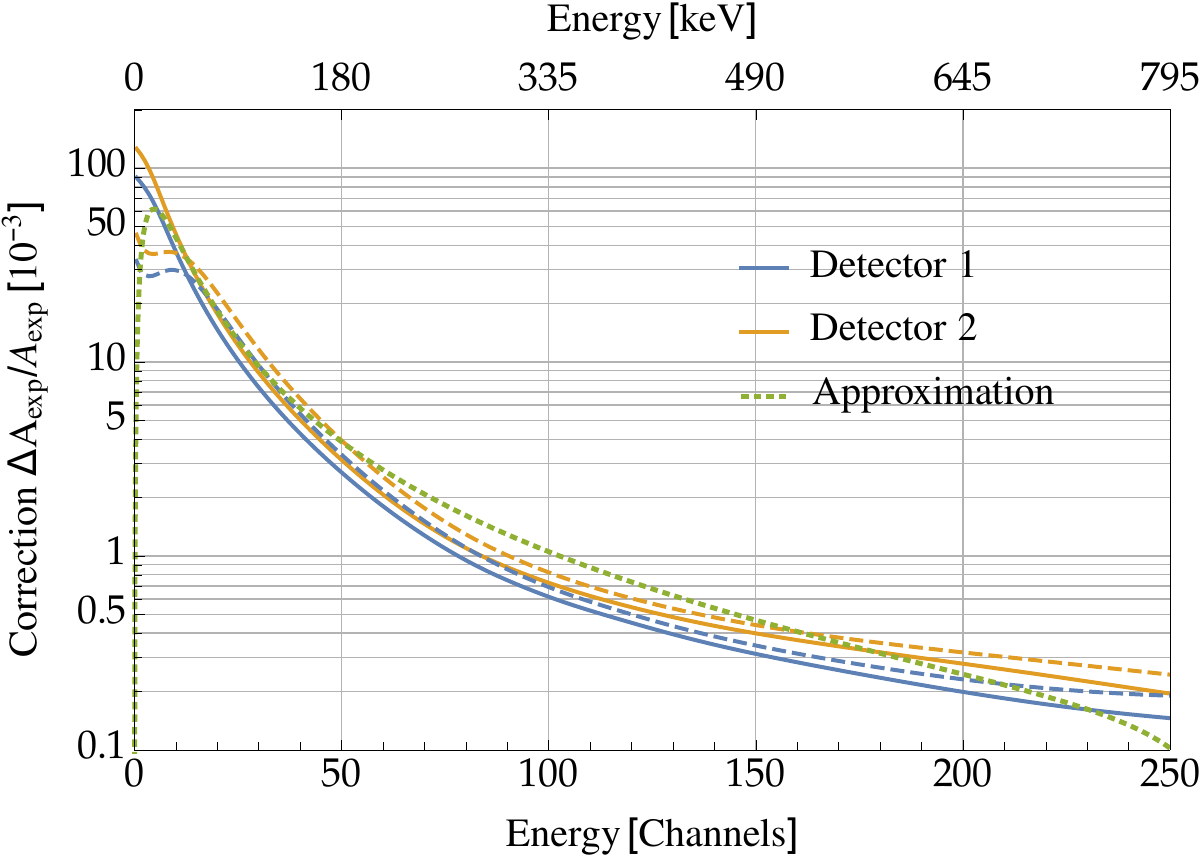}
\caption{Undetected backscattering to the opposite detector decreases the measured asymmetry and therefore leads to a relative correction with positive sign. Solid lines show the correction without energy reconstruction, dashed lines show the correction when the energy information of the undetected impact is not lost. The dotted line shows an estimation with very simple assumptions about the detector response, e.g. without non-linearity and a simplified trigger function.}
\label{fig:ubscorrection}
\end{figure}

The solid lines in \figref{fig:ubscorrection} show the resulting corrections to $A_\textup{exp}$ measured with \perkeo{} according to eq. \eqref{eq:asymcorrected}. For the determination of $\lambda$, only events with a detected energy $>\SI{300}{keV}$ are considered, which results in an overall fractional correction of \num{1.3e-4}, which corresponds to a correction of \num{5e-4} on $A$.

Table \ref{tab:uncertainties} summarizes all uncertainties to an overall relative uncertainty of the correction of \SI{27}{\percent}.
\begin{table}[h]
\centering
\caption{Overview of systematic uncertainties in the correction.}
 \begin{tabular}{l|c}
    systematic effect & magnitude\\
    \hline
    trigger function & \SI{5}{\percent}\\
    magnetic field  & \SI{10}{\percent}\\
    backscattering coefficients & \SI{20}{\percent}\\
    energy reconstruction & \SI{15}{\percent}\\
    \hline
    combined & \SI{27}{\percent}
 \end{tabular}
 \label{tab:uncertainties}
\end{table}

A similar effect, with different materials however, also plays a role in the UCNA experiment, where ultra-cold neutrons are used to determine the electron asymmetry parameter. While the overall correction due to this effect is much larger, the relative uncertainty is determined to be of the same magnitude \cite{UCNACollaborationPrecisionmeasurementneutron2013}.

\subsection{Influence on the Beta Decay Spectrum}
An analysis of the unpolarized neutron beta decay spectrum enables tests of the Standard Model by looking for a non-zero Fierz interference term \cite{Gonzalez-AlonsoKinematicsensitivityFierz2016, HickersonFirstdirectconstraints2017}, which would hint to scalar or tensor interactions. Undetected backscattering imposes a systematic correction by shifting the measured spectrum towards lower energies. An analysis of \perkeo{} data starting at channel \num{50} (which corresponds to \SI{180}{keV}) would lead to a shift of $\approx\num{-2.6e-3}$ of the measured Fierz interference term $b$, if undetected backscattering without complete energy reconstruction were missed. If the energy can be fully reconstructed, only the decreased trigger probability due to backscattering leads to a distortion of the spectrum and would have negligible influence of $\sim\num{1e-4}$.

\section{Summary}
We show that undetected backscattering from scintillation detectors can have a sizable effect on the measurement of neutron beta decay parameters. In order to determine a correction, the knowledge about the energy dependent trigger function of the detector and backscattering probabilities is essential. We determine a correction to the experimental electron asymmetry measured with \perkeo{} with a conservative uncertainty of \SI{27}{\percent}. A determination of the Fierz interference term from the unpolarized neutron decay spectrum may be significantly affected by the effect.

\begin{acknowledgement}
This work was supported by the Priority Programme SPP~1491 of the German Research Foundation (DFG), contract nos. MA~4944/1-2 and AB~128/5-2, the Austrian Science Fund (FWF) contract nos. P~26636-N20 and I~534-N20, the German Federal Ministry for Research and Education, contract nos. 06HD153I and 06HD187, and the cluster of excellence `Origin and Structure of the Universe' of the German Research Foundation.
\end{acknowledgement}

%

\end{document}